\documentclass[aps,prb,reprint,superscriptaddress,floatfix]{revtex4-2}

\usepackage{dcolumn}
\usepackage{bm}
\usepackage{amsmath}
\usepackage[percent]{overpic}
\usepackage{mathrsfs}

\newcommand{\inst}[1]{\affiliation{#1}}
\newcommand{\abst}[1]{\begin{abstract}#1\end{abstract}}
\begin{document}

\title{Dynamical scaling method improved by a deep learning approach}
 

\author{Yusuke Terasawa and Yukiyasu Ozeki}
\inst{%
Department of Engineering Science, Graduate School of Informatics and Engineering, The University of Electro-Communications, 1-5-1 Chofugaoka, Chofu-shi, Tokyo 182-8585, Japan
}

\abst{
We propose a dynamical scaling analysis improved by a deep learning approach. 
While Gaussian process regression has been widely employed for estimating scaling parameters, its computational cost for parameter optimization becomes a limitation in dynamical scaling analysis, where large datasets are involved.
In contrast, the present method employs a neural network, which significantly reduces the computational cost and enables the use of the entire dataset that was inaccessible with Gaussian process regression. 
We applied the method to the 2D Ising model and the 2D 3-state Potts model, achieving higher accuracy and computational efficiency than conventional approaches.
}
\maketitle
\section{Introduction}

Critical phenomena are ubiquitous in nature and are observed across a wide range of systems in physics, biology, and information science. These phenomena are characterized by scale invariance and scaling laws that emerge near critical points. Understanding critical phenomena is not only essential for elucidating the fundamental mechanisms of phase transitions but also has significant implications for applications in diverse fields. A central objective in the study of critical phenomena is the classification of systems into universality classes based on their shared critical exponents. Accurate determination of universality classes and scaling behavior remains a crucial challenge for both theoretical advancement and practical application.

The most widely used technique for analyzing critical behavior in spin systems is the finite-size scaling (FSS) method\cite{cardy1988}. While this method has been shown to be a powerful tool with numerous successful applications, it requires the system to be equilibrated before analysis. As a result, the computational cost becomes prohibitively high, particularly for systems with slow relaxation dynamics.

In such cases, the nonequilibrium relaxation (NER) method \cite{Ozeki2007} provides an effective alternative. This approach investigates the critical behavior by analyzing the relaxation process from an initial state toward equilibrium.  
An additional advantage of NER analysis is that it provides access to a significantly larger amount of data compared to FSS, which can be effectively used in the NER analysis.
These advantages make the NER method particularly suitable for studying systems exhibiting slow relaxation, such as frustrated or disordered systems\cite{ozeki2014,ozeki2012,ozeki2009,Yamamoto2004,Nakamura2006,Ozeki2003,ito2003}.

The scaling technique employed in the NER method is known as dynamical scaling analysis.  
This approach has been shown to be effective, particularly in estimating transition temperatures in Kosterlitz-Thouless (KT) transitions~\cite{Ozeki2003} and critical exponents in spin-glass transitions~\cite{Yamamoto2004,Nakamura2006,Terasawa2023}.  
However, it requires a parametric model function for the fitting procedure, which introduces issues related to model-dependent uncertainties.  
To solve this problem, a scaling analysis method based on Gaussian process regression (GPR) has been proposed for dynamical scaling~\cite{Echinaka2016}, which was originally developed in the context of FSS analysis~\cite{Harada2011}.
This GPR-based method enables high-precision analysis without assuming any specific model function, thereby eliminating the associated systematic errors~\cite{Echinaka2016,Ozeki2020,Murayama2020,Nakamura2016,Nakamura2019,Nakamura2025}.

However, a major drawback of GPR is its computational complexity of $O(N^3)$, where $N$ is the number of data points. Since dynamical scaling analysis typically involves a much larger number of data points than FSS, applying GPR in this context often requires discarding a significant portion of the data, which may lead to deviations in the estimated results. Overcoming this issue requires the development of more computationally efficient estimation methods.

Recently, it has been demonstrated that applying deep learning to FSS analysis can reduce the computational cost from $O(N^3)$ to $O(N)$ \cite{Yoneda2023}. This suggests that incorporating deep learning techniques into dynamical scaling analysis has the potential to overcome the aforementioned limitations and enable more accurate and efficient characterization of critical behavior.

The objective of the present study is to validate a dynamical scaling analysis method combined with neural networks.
In the proposed analysis, the transition temperature is treated as a key parameter and serves as a primary criterion for validation.
Therefore, validation requires a benchmark system whose critical temperature is known exactly, enabling a direct and quantitative assessment of the inferred transition temperature.
For this purpose, we consider the two-dimensional Ising model and the two-dimensional three-state Potts model, both of which have exactly known critical temperatures. 
We then compare the resulting estimates with those obtained by GPR, thereby establishing the reliability of the present method and assessing its applicability to a broader class of systems.

The structure of this paper is as follows. In Sec.~\ref{sec:Method}, we describe the models and observables and formulate the dynamical scaling law. We also present the details of the neural network approach and the GPR used for comparison. In Sec.~\ref{result:2D ising model}, we apply the proposed method to the 2D Ising model and examine its accuracy and stability. In Sec.~\ref{sec:Result 2D 3state potts}, we further perform the same analysis for the 3-state Potts model and evaluate the performance of the method. Finally, a summary and future perspectives are given in Sec.~\ref{sec:Summary}.

\section{Method}
\label{sec:Method}
\subsection{Model, observables, and dynamical scaling law}
In this study, we consider the 2D Ising model and the 2D 3-state Potts model. The Hamiltonian of the 2D Ising model is given by
\begin{equation}
H_{\mathrm{Ising}} = -J \sum_{\langle i,j\rangle} S_i S_j ,
\end{equation}
where $S_i = \pm 1$. The Hamiltonian of the 2D 3-state Potts model is defined as
\begin{equation}
H_{\mathrm{Potts}} = -J \sum_{\langle i,j\rangle} \delta_{S_i, S_j} ,
\end{equation}
where $S_i \in {1,2,3}$ and $\delta_{S_i,S_j}$ denotes the Kronecker delta.

In both cases, the summation $\sum_{\langle i,j\rangle}$ is taken over all nearest-neighbor pairs. The temperature $T$ is scaled by $J/{k_\mathrm{B}}$. The linear system size is denoted by $L$. The total number of spins $N_{\mathrm{spins}}$ is $L(L-1)$, and skewed periodic boundary condition is imposed. \par
In dynamical scaling analysis, we first compute the relaxation process of magnetization starting from the all-aligned state as the initial configuration. Let $t$ denote the number of Monte Carlo steps (MCS) elapsed from the initial state, and let $T$ be the temperature of the system. For the 2D Ising model, the magnetization per spin at time $t$ and temperature $T$ is defined as
\begin{equation}
m(t,T)=\frac{1}{N_{\mathrm{spins}}}\sum_i S_i(t,T),
\end{equation}
where the initial condition is taken as $S_i(0,T)=+1$ for all $i$.
For the 2D 3-state Potts model, we set the initial condition to $S_i(0,T)=1$ for all $i$. The magnetization is then defined as
\begin{equation}
m(t,T)= \frac{3}{2}\left(\frac{1}{N_{\mathrm{spins}}}\sum_i \delta_{S_i(t,T),1}-\frac{1}{3}\right).
\end{equation}

It is expected that a dynamical scaling law holds in the asymptotic regime:
\begin{equation}
\label{mag scaling form}
m(t,T) = t^{-\lambda} \Phi(t/\tau(T)),
\end{equation}
where $\lambda$ denotes a local exponent of magnetization and $\tau(T)$ represents the relaxation time.
The functional form of $\tau(T)$ depends on the type of the phase transition.  
For a second-order transition, the following asymptotic behavior is expected:
\begin{equation}
    \tau(T) \sim |T - T_{\mathrm{c}}|^{-b},
\end{equation}
where \( b \) is a tunable parameter to be optimized, and \( T_{\mathrm{c}} \) denotes the critical temperature of the second-order transition.  
Taking this form into account, the scaling relation in Eq.~(\ref{mag scaling form}) can be formed as
\begin{equation}
    \label{mag scaling form by XY}
    Y = \Phi(X),
\end{equation}
where
\begin{equation}
    \label{Y scaling form}
    Y = t^\lambda m(t,T),
\end{equation}
\begin{equation}
    \label{X scaling form}
    X = t / \tau(T).
\end{equation}

The expressions in Eqs.~(\ref{Y scaling form}) and (\ref{X scaling form}) contain three physical parameters to be optimized: \( (T_{\mathrm{c}} ,\, \lambda,\, b) \).  
The scaling relation given in Eq.~(\ref{mag scaling form by XY}) asserts that, if these parameters are correctly chosen, the data points \((X, Y)\)  collapse onto a single scaling function.\par
In the present study, we focus on the estimation of the critical temperature \( T_{\mathrm{c}} \) as a key parameter. This is because a correct estimation of critical exponents requires taking finite-time corrections into account, whereas Eq.~(\ref{mag scaling form}) does not include such corrections. 
A similar issue is also present in a previous study based on FSS\cite{Yoneda2023}.

\subsection{Neural Networks}
\label{sec:Neural Networks}

\subsubsection{General expression of neural networks}
Let us describe how the transition temperature can be estimated by combining the dynamical scaling law with neural networks. In this approach, we construct the scaling function $\Phi(\cdot)$, as expressed in Eq.~(\ref{mag scaling form by XY}), using a neural network. 
By optimizing the physical parameters involved in the scaling relations, such as the transition temperature, simultaneously with the internal parameters of the neural network, we fit the transformed data $(X, Y)$ that follow Eqs.~(\ref{Y scaling form}) and (\ref{X scaling form}), following the method proposed in Ref.~\cite{Yoneda2023}. In the present study, we apply this approach to the context of dynamical scaling analysis. \par
\begin{figure}[htbp]
    \centering
    \includegraphics[width=1.0\linewidth]{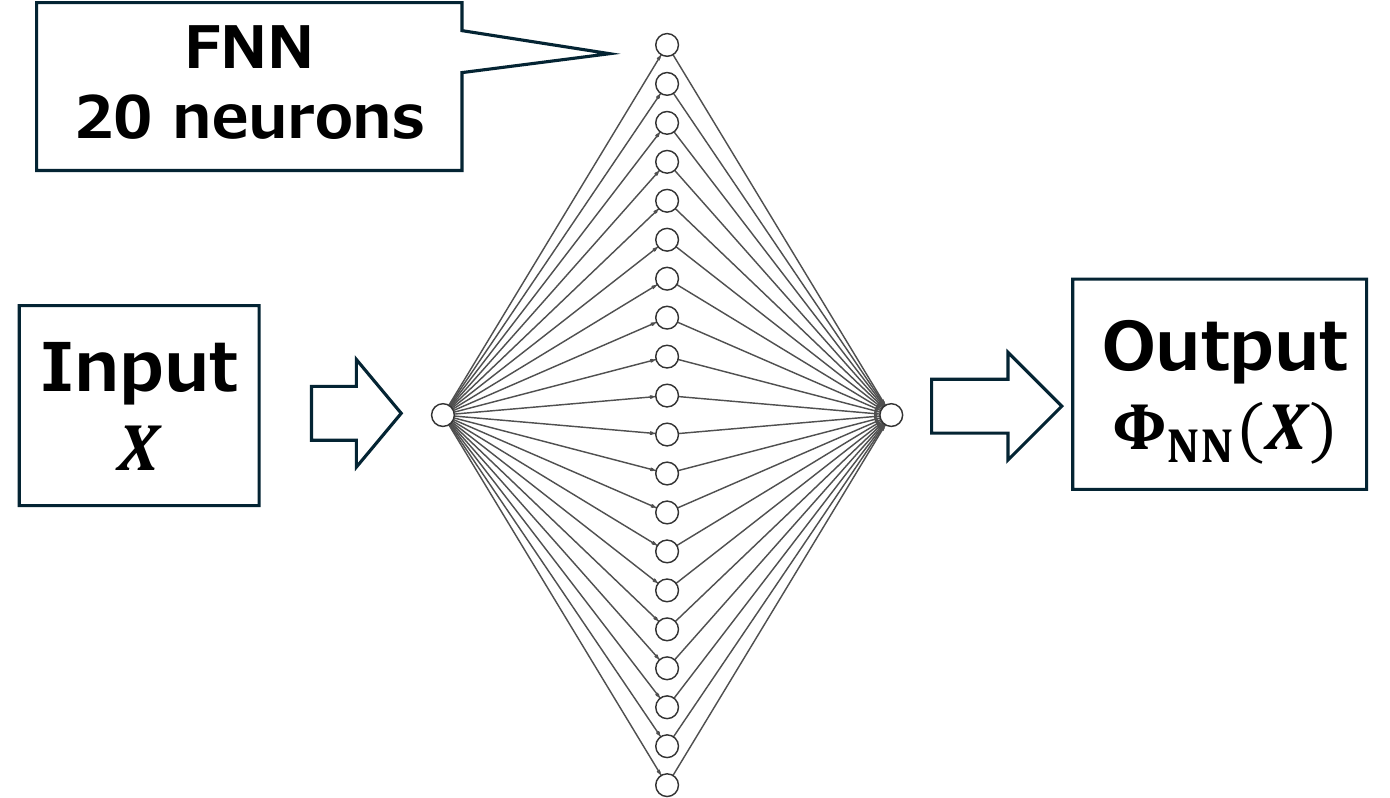}
    \caption{The architecture of the neural network used in this study. FNN denotes a fully connected neural network. The input and output layers consist of a single neuron, while the hidden layer contains 20 neurons. The activation function is the softplus function.}
    \label{fig:NN arch}
\end{figure}

We employ a fully connected neural network to represent the dynamical scaling function. The network consists of 3 layers: an input layer, 1 hidden layer, and an output layer. This hidden layer contains 20 neurons, and both the input and output layers have a single neuron, reflecting the one-dimensional nature of the input and output variables. This network architecture is based on the one employed in a previous study~\cite{Yoneda2023}, 
but we reduce the number of neurons in the hidden layer to construct a smaller network for the present analysis. 
This modification is introduced to reduce the computational cost, and we have confirmed that the resulting accuracy remains comparable to that of the original architecture.
The structure is illustrated in Fig.~\ref{fig:NN arch} The weights and biases are initialized independently using a zero-mean uniform distribution $U(-1/\sqrt{n},\,1/\sqrt{n})$, where $n$ is the number of neurons in the preceding layer. This initialization is motivated by standard practices for deep neural networks~\cite{He2015ICCV} and facilitates stable optimization.

All layers are connected via affine transformations followed by elementwise application of a nonlinear activation function. In the present study, we employ the softplus function, which is defined as
\begin{equation}
    \varphi(x)=\ln{\{1+\exp(x)\}}.
\end{equation}
The model parameters are optimized so as to minimize a loss function, which is measured by the mean squared error and defined as
\begin{equation}
    \mathcal{L}_{\mathrm{NN}} = \frac{1}{N_{\mathrm{data}}}\sum_{i=1}^{N_{\mathrm{data}}} \bigg(Y_i - \Phi_{\mathrm{NN}}(X_i)\bigg)^2.
\end{equation}
Here, \(\Phi_{\mathrm{NN}}(\cdot)\) denotes the dynamical scaling function represented by the neural network. \(\mathcal{L}_{\mathrm{NN}}\) is minimized with respect to both the physical parameters \((T_{\mathrm{c}}, \lambda, b)\) and the internal parameters of the neural network. The computational cost of a gradient step scales as \(O(N)\).\par
For the optimization algorithm, we employ the Adam optimizer~\cite{kingma2017Adam} with a learning rate of \(1 \times 10^{-3}\), as recommended in the original paper.
To improve the accuracy and stability of the parameter estimation, we adopt the minibatch learning scheme~\cite{minibatch}, which was not employed in previous study\cite{Yoneda2023}. In minibatch learning, the entire dataset is randomly divided into small subsets, each of which contains a fixed number of data points referred to as a minibatch. For each minibatch, the gradient of the loss function is computed and used to update the model parameters. Once all minibatches have been used for parameter updates, one epoch is said to have been completed. By repeating this process over many epochs, the minimum of the loss function is approximately evaluated. In the present study, double-precision floating-point arithmetic was employed to ensure higher numerical accuracy.

\subsubsection{Data preprocessing}
In order to obtain good performance in machine learning, it is essential to transform the data into a form that facilitates fitting. 
In the dynamical scaling analysis of second-order transitions, the relaxation time $\tau(T)$ exhibits a characteristic critical behavior, and the data must be transformed accordingly. 
For such transitions, we apply the following transformation:
\begin{equation}
    T \mapsto \frac{T - T_{\mathrm{min}}}{T_{\mathrm{max}} - T_{\mathrm{min}}}
\end{equation}
\begin{equation}
    t \mapsto \frac{t}{t_{\mathrm{max}}}
\end{equation}
\begin{equation}
    m(t,T) \mapsto \frac{m(t,T)}{m_{\mathrm{max}}}
\end{equation}
Here, $t_{\mathrm{max}}$ and $m_{\mathrm{max}}$ denote the maximum values of $t$ and $m(t,T)$ in the dataset under analysis, respectively. $T_{\mathrm{max}}$ and $T_{\mathrm{min}}$ represent the maximum and minimum temperatures in the dataset, respectively. These transformations, motivated in part by preprocessing approaches found in previous studies\cite{Yoneda2023}, normalize the dataset $\{t, T, m\}$ into the range $[0,1]$, which facilitates stable and accurate machine learning estimations in dynamical scaling analysis.

\subsection{Gaussian Process Regression}
\label{method:GPR}
To compare with the results obtained using the method described in Sec.~\ref{sec:Neural Networks}, we briefly explain the dynamical scaling analysis based on GPR~\cite{Echinaka2016}. Let $\vec{X}$ and $\vec{Y}$ denote the vectors formed by the scaled variables $X$ and $Y$ in Eq. (\ref{mag scaling form by XY}), respectively. The corresponding dynamical scaling function evaluated at $\vec{X}$ is denoted by $\vec{\Phi}$. The log-likelihood function $\ln{\mathcal{L}_{\mathrm{GPR}}(\vec{\theta})}$, given a parameter set $\vec{\theta}$ that includes both physical and hyper parameters, can then be written as
\begin{equation}
    \label{GPR Loglikelihood}
    \ln{\mathcal{L}_{\mathrm{GPR}}(\vec{\theta})} = -\frac{1}{2}\ln{|2\pi\Sigma|} - \frac{1}{2}(\vec{Y} - \vec{\Phi})^t \Sigma^{-1} (\vec{Y} - \vec{\Phi}),
\end{equation}
where $\Sigma$ is the covariance matrix and its elements $(i,j)$ defined via the kernel function $K(X_i,X_j,\vec{\theta})$ as
\begin{equation}
    (\Sigma)_{ij} = E_i^2\delta_{ij} + K(X_i, X_j, \vec{\theta}),
\end{equation}
\begin{equation}
    K(X_i, X_j, \vec{\theta}) = \theta_0^2\delta_{ij} + \theta_1^2 \exp\left(-\frac{(X_i - X_j)^2}{2\theta_2^2}\right).
\end{equation}
Here, $E_i$ denotes the error associated with $Y_i$, and $(\theta_0, \theta_1, \theta_2)$ are the hyperparameters of the kernel function. In practice, the following logarithmic transformations were applied to the dataset as a preprocessing step:
\begin{align}
X_i^\prime &\equiv \ln X_i , \notag \\
Y_i^\prime &\equiv \ln Y_i , \notag \\
E_i^\prime &\equiv \ln \left( 1 + \frac{E_i}{Y_i} \right) \approx \frac{E_i}{Y_i} .
\end{align}
Since the evaluation of Eq.~(\ref{GPR Loglikelihood}) requires $O(N^3)$ computational cost in general, primarily due to the inversion of the covariance matrix, it is difficult to perform inference using the entire dataset. Therefore, 100 points per temperature were extracted, evenly spaced in $\ln t$, excluding the initial relaxation regime, to construct the dataset for GPR analysis.

\section{Result}
\label{sec:Result}
\subsection{2D Ising model}
\label{result:2D ising model}

To verify the validity and computational efficiency of the proposed method, we first applied the dynamical scaling analysis to the two-dimensional Ising model, for which the exact value of the critical temperature is known.
The data used in the analysis were generated by Markov chain Monte Carlo (MCMC) simulations employing the Metropolis update scheme, starting from the all-aligned state. This simulation was performed with a system size of $L = 20001$, up to $10^5$ MCS, and averaged over 100 samples.

\begin{figure}[htbp]
    \centering
    \begin{overpic}[width=1.0\linewidth]{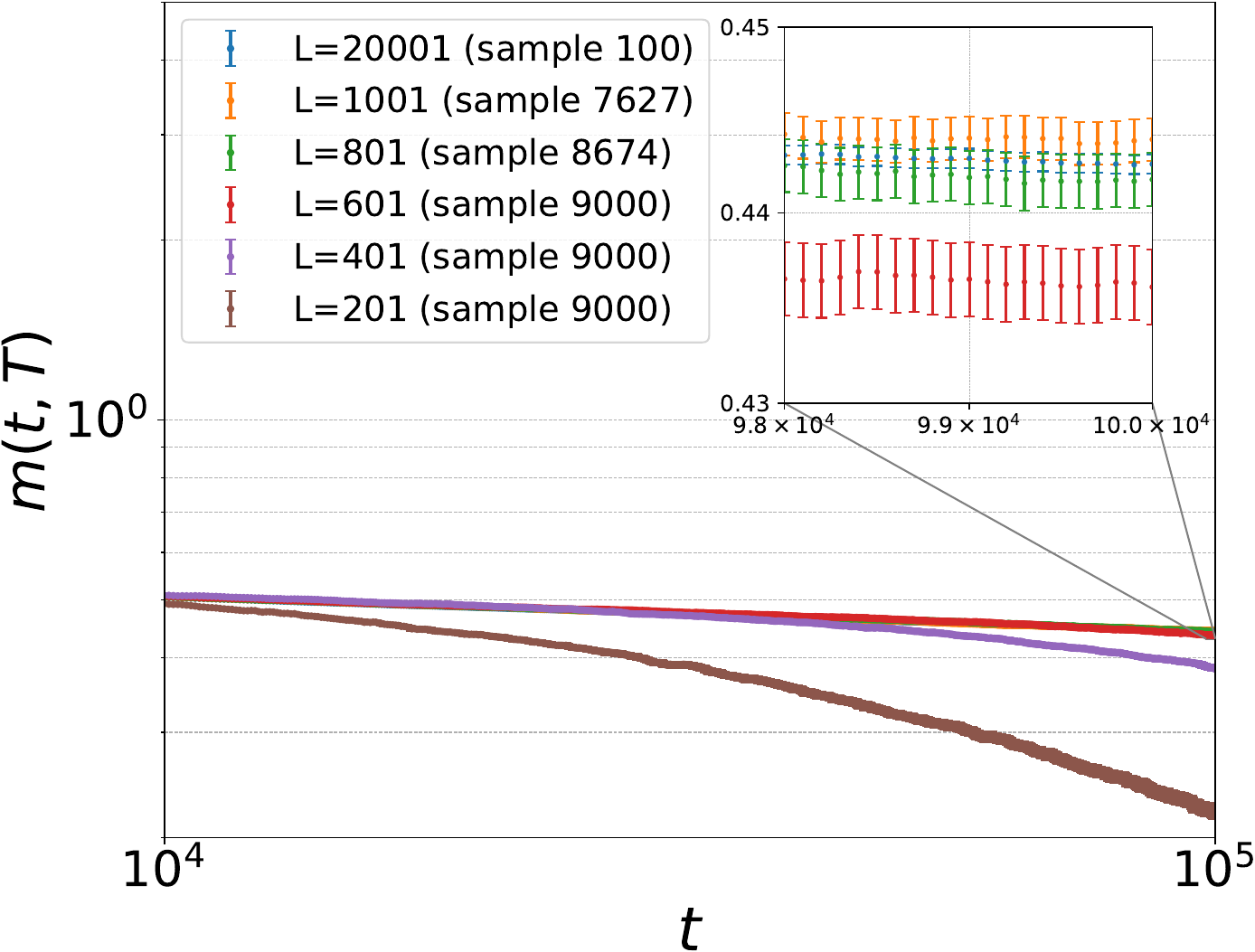}
    \end{overpic}
    
    \caption{Size dependence for the 2D Ising model. It shows that the dependence becomes negligible for $L \geq 1001$. The data are plotted every 100 MCS. The legends indicate the number of samples used in each simulation. 
    }
    \label{fig:size dependence}
\end{figure}

\begin{figure}[htbp]
    \centering
    \begin{overpic}[width=1.0\linewidth]{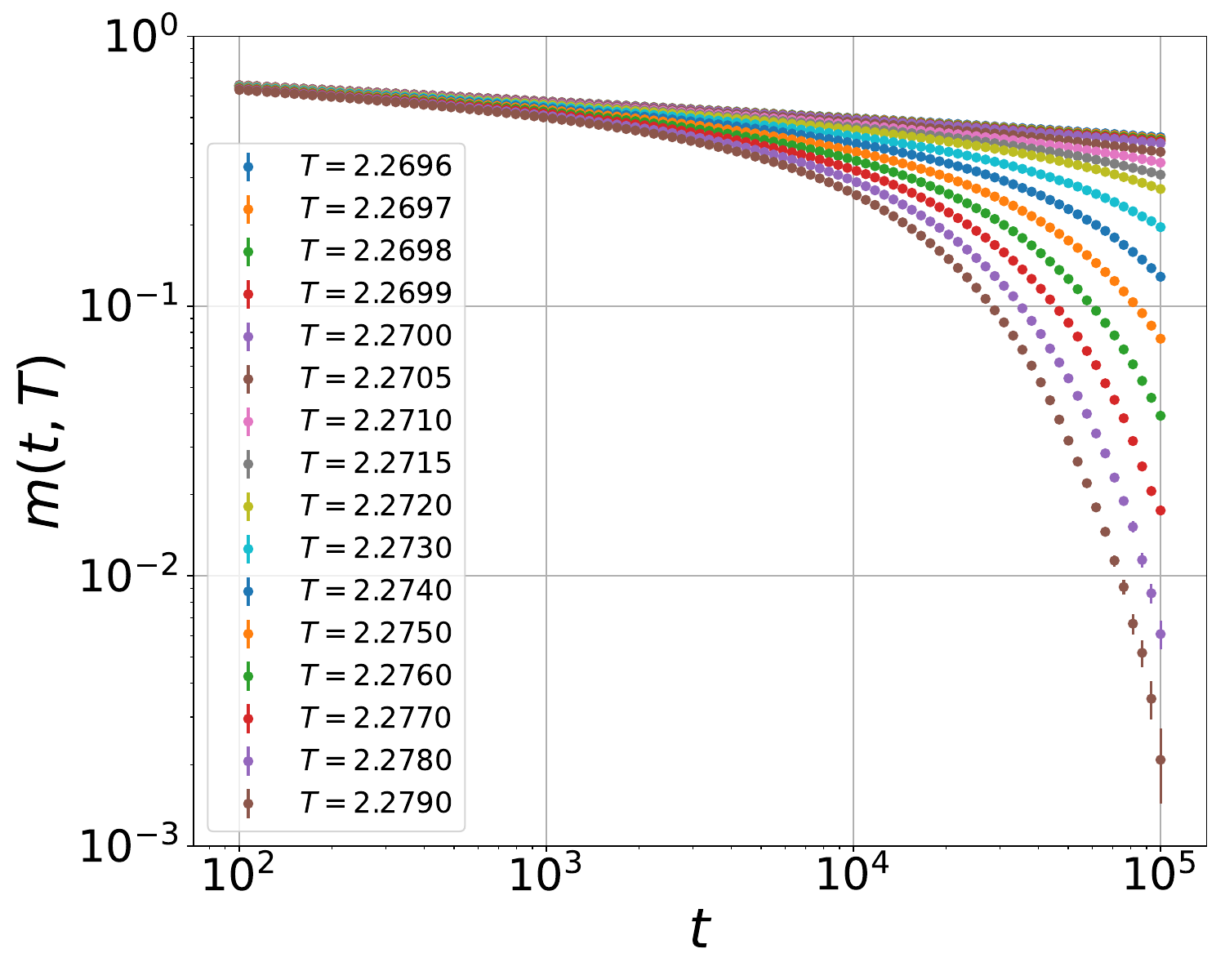}
        \put(90,80){\large (a)} 
    \end{overpic}
    \vspace{0.5em}
    \begin{overpic}[width=1.0\linewidth]{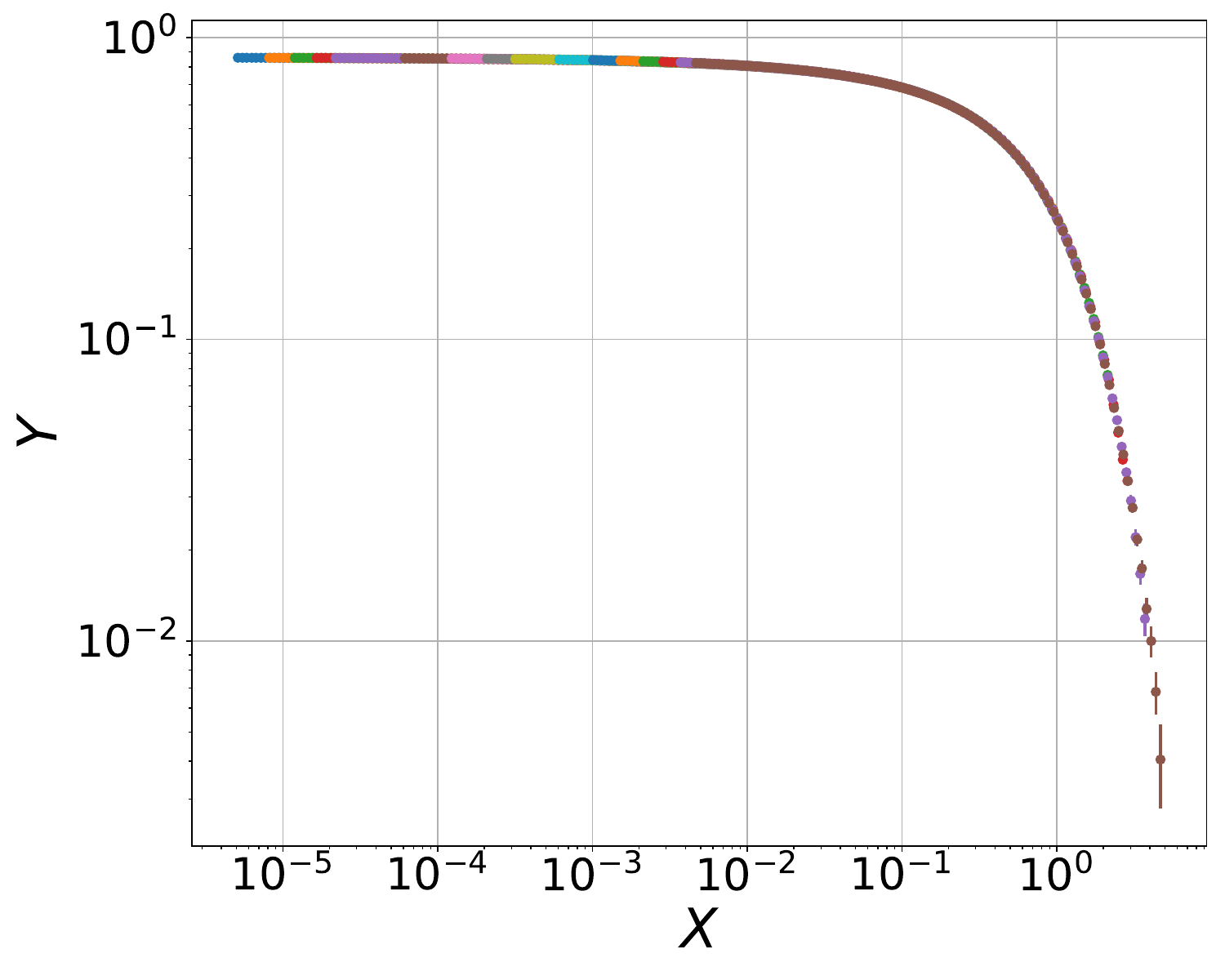}
        \put(90,80){\large (b)} 
    \end{overpic}
    \caption{Relaxation data and dynamical scaling plot for the 2D Ising model used in the analysis. Both panels are plotted with 101 data points per temperature for visualization. (a) shows the relaxation data after discarding the initial relaxation regime, which was used for parameter estimation using a neural network. (b) shows the corresponding scaling plot. The estimated parameters are $T_{\mathrm{c}} = 2.269186(1)$, $\lambda = 0.0577(1)$, and $b = 2.1555(3)$.}
    \label{fig:2D ising model relaxation and scaling}
\end{figure}

For a reliable scaling analysis, it is essential that the data be free from significant finite-size effects. As shown in Fig.~\ref{fig:size dependence}, the system size used in the simulation is large enough to ensure that finite-size effects are negligible. The resulting relaxation data are shown in Fig.~\ref{fig:2D ising model relaxation and scaling}.  Based on the relaxation data, a dataset for scaling analysis was prepared by excluding the initial relaxation regime corresponding to $t < 100$ MCS. The total number of data points in this dataset was 1,598,416.\par

We next describe the determination of the parameters required for the dynamical scaling analysis using a neural network, as well as the results obtained. 
The initialization of the model architecture, weights, and biases was performed as described in Sec.~\ref{sec:Neural Networks}. 
To determine an appropriate batch size, we examined how the estimated critical temperature depends on the choice of batch size. 
To this end, we performed calculations while keeping the total computational time approximately the same, and varied the batch size in powers of two. 
During these runs, we monitored the optimization by tracking the loss function and confirmed that it reaches a stable plateau.

These error bars were estimated using the bootstrap method~\cite{Efron1979} with 128 resamplings. Specifically, for each resampling, a new dataset was generated by sampling with replacement from the original dataset, and an independent optimization was carried out. The final estimates were taken as the mean of the optimized parameters obtained from the resampled datasets, and their uncertainties were evaluated as the standard deviation, corresponding to the $1\sigma$ confidence level.

\begin{figure}[htbp]
    \centering
    \includegraphics[width=1.0\linewidth]{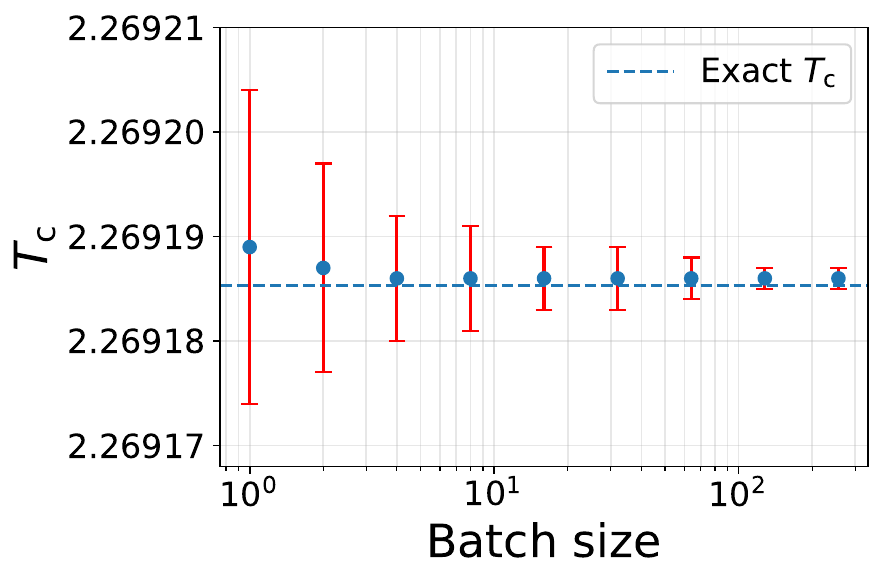}
    \caption{Estimated values of $T_{\mathrm{c}}$ as a function of batch size in the optimization of dynamical scaling parameters for the 2D Ising model. The horizontal blue line indicates the exact solution for the critical temperature. The error bars were estimated using the bootstrap method with 128 resamplings}
    \label{fig:2d ising batchsize dependence}
\end{figure}

\begin{figure}[htbp]
    \centering
    \includegraphics[width=1.0\linewidth]{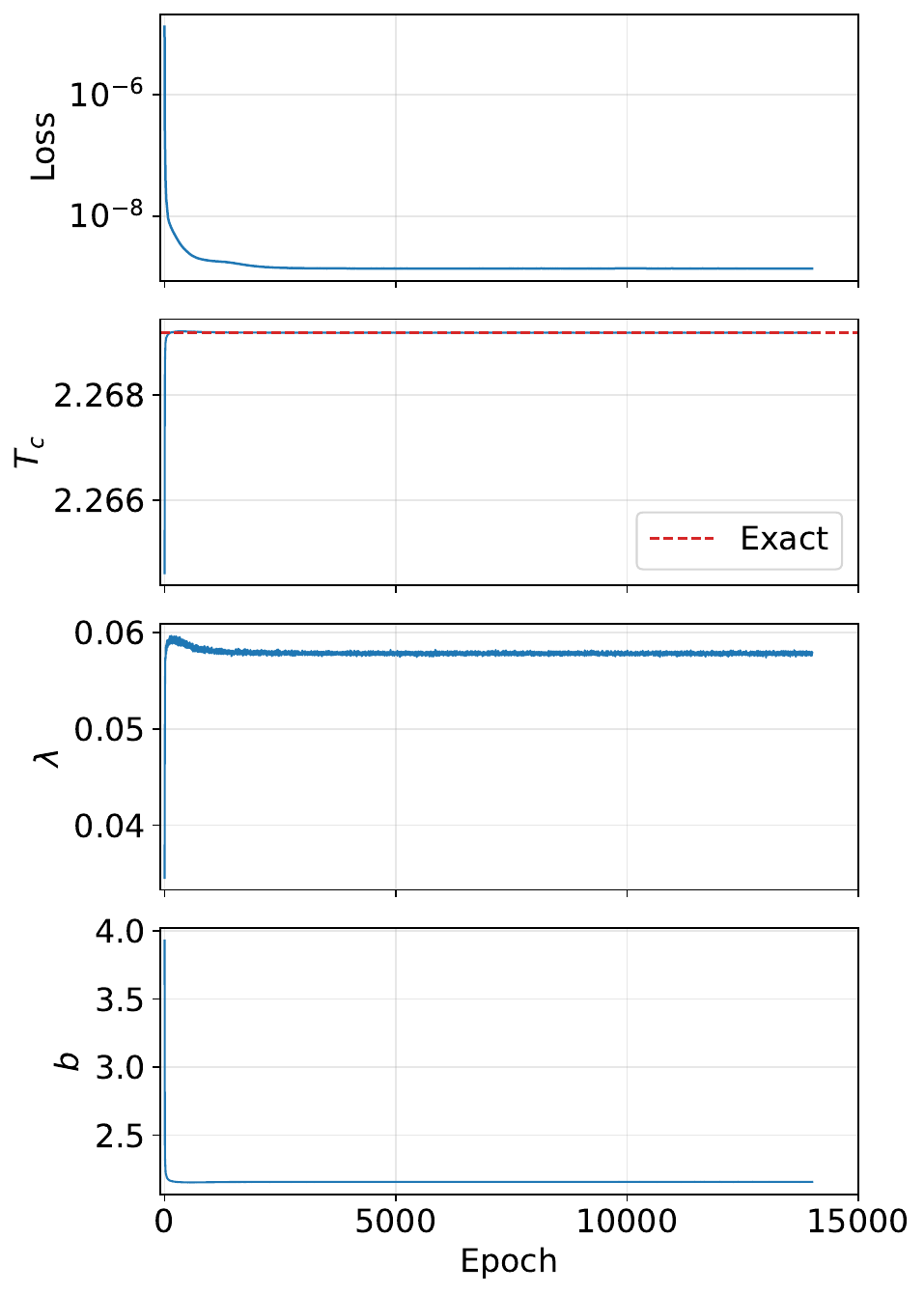}
    \caption{Optimization process of the dynamical scaling parameters in the 2D Ising model at batch size $256$. Each panel shows the optimization history of an individual parameter. The horizontal axis represents the number of epochs and is shared across all panels. In the second panel, which corresponds to the critical temperature $T_{\mathrm{c}}$, the exact solution is indicated by a reference line.}
    \label{fig:2d Ising model parameters 2x2}
\end{figure}

The results are shown in Fig.~\ref{fig:2d ising batchsize dependence} and  Table \ref{table:batchsize dependence}. We find that increasing the batch size tends to improve the estimation accuracy. However, the improvement saturates beyond a certain batch size. Based on these observations, we adopt the results obtained with the largest batch size feasible in our computations, $256$, as the final estimates in this study. Figure~\ref{fig:2d Ising model parameters 2x2} shows the evolution of the inferred physical parameters during the optimization process at this batch size, indicating that the estimates are sufficiently converged.

The exact value, the final estimated value, and the result obtained from GPR are summarized in Table~\ref{2d ising result table}. The dataset used for the GPR analysis was constructed as described in Sec.~\ref{method:GPR}, and consisted of a total of 1600 data points.

The result of the dynamical scaling analysis using the neural network was found to be consistent with the exact solution of critical temperature. This confirms the validity of the present method. Note that the critical exponents are not discussed here because finite-time corrections have not been taken into account in this method. An approach to incorporate such corrections is to perform an extrapolation of the nonequilibrium relaxation of fluctuations just above the critical temperature\cite{Osada2024}.

\begin{figure}[htbp]
    \centering
    \begin{overpic}[width=1.0\linewidth]{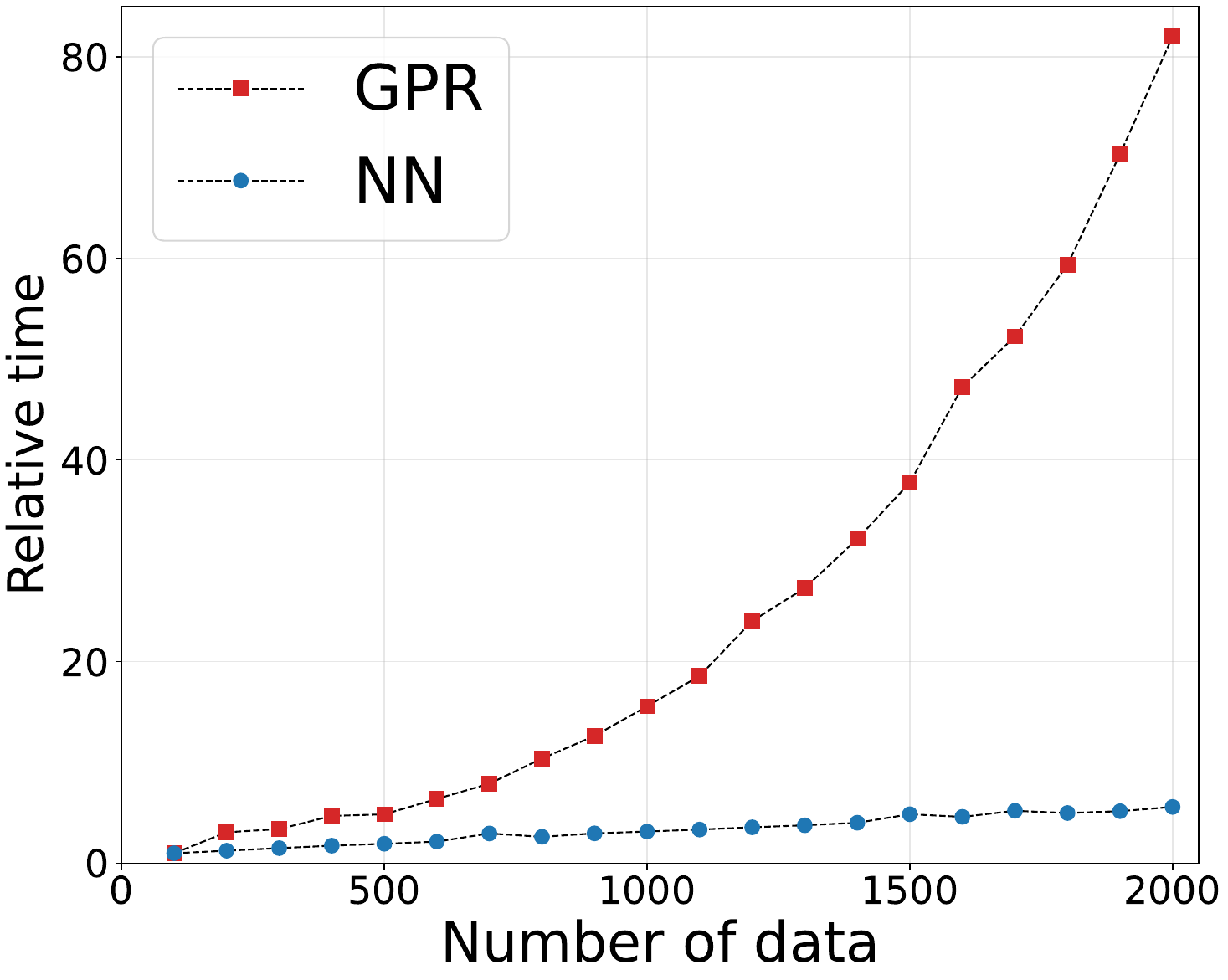}
    \end{overpic}
    
    \caption{Relative computation time required for one epoch update for GPR and NN. The computation time for 100 data points is taken as the unit of time. The dataset consists of NER data for the 2D Ising model.
    }
    \label{fig:GPR vs NN computational time}
\end{figure}

As a supplementary analysis, we compare the computational cost of GPR and NN. Specifically, we first prepare a dataset in the temperature range $2.270 \le T \le 2.279$ with a temperature interval of $\Delta T = 0.001$. The simulation conditions are the same as those described above. We then measure the computation time required for one epoch for both GPR and NN while varying the number of data points from 100 to 2000 in increments of 100. The data points are selected at equal intervals on a logarithmic scale. As the unit of time, we use the relative computation time normalized independently for each method by the computation time at 100 data points. The reported values are the averages over 100 independent runs. We note that a similar measurement was also performed in a previous study \cite{Yoneda2023}. The obtained results are shown in Fig.~\ref{fig:GPR vs NN computational time}. These results confirm that the NN-based analysis substantially reduces the computational cost compared with the GPR-based analysis. This advantage makes it possible to fully utilize the large amount of data obtained from NER simulation.

\begin{table}
\centering
\caption{Comparison of the estimated scaling parameters for the 2D Ising model.}
\label{2d ising result table}
\setlength{\tabcolsep}{4pt} 
\begin{tabular}{clll}\hline\hline
Reference &  $T_{\mathrm{c}}$ & $\lambda$ & $b$ \\ \hline\hline
Exact Value & 2.26918531$\ldots$ & & \\ \hline
Neural Network & 2.269186(1) & 0.0577(1) & 2.1555(3) \\
GPR & 2.269238(4) & 0.05839(2) & 2.139(2) \\
\hline\hline
\end{tabular}
\end{table}

\begin{table*}[t]
\centering
\caption{Dependence of the estimated parameters on the batch size in the 2D Ising model. The number of epochs is the effective number achieved under a fixed computational budget.}
\label{table:batchsize dependence}
\begin{tabular}{c c l l l}\hline\hline
Batch size & Used epochs & $T_c$ & $\lambda$ & $b$ \\
\hline\hline
1   &   125  & 2.269189(15) & 0.0578(11) & 2.154(4) \\
2   &   232  & 2.269187(10) & 0.0579(9)  & 2.155(2) \\
4   &   484  & 2.269186(6)  & 0.0578(7)  & 2.155(2) \\
8   &   944  & 2.269186(5)  & 0.0578(3)  & 2.155(1) \\
16  &  1820  & 2.269186(3)  & 0.0578(2)  & 2.1556(9) \\
32  &  3359  & 2.269186(3)  & 0.0577(2)  & 2.1554(7) \\
64  &  5948  & 2.269186(2)  & 0.0577(2)  & 2.1555(5) \\
128 &  9134  & 2.269186(1)  & 0.0577(1)  & 2.1555(3) \\
256 & 14008  & 2.269186(1)  & 0.0577(1)  & 2.1555(3) \\ \hline\hline
\end{tabular}
\end{table*}

\subsection{2D 3-state Potts model}
\label{sec:Result 2D 3state potts}
To further assess the performance of the proposed method, we carried out the same analysis for the 2D 3-state Potts model. The data were generated by MCMC simulations using the Metropolis update scheme, in the same manner as for the 2D Ising model.

\begin{figure}[htbp]
    \centering
    \begin{overpic}[width=1.0\linewidth]{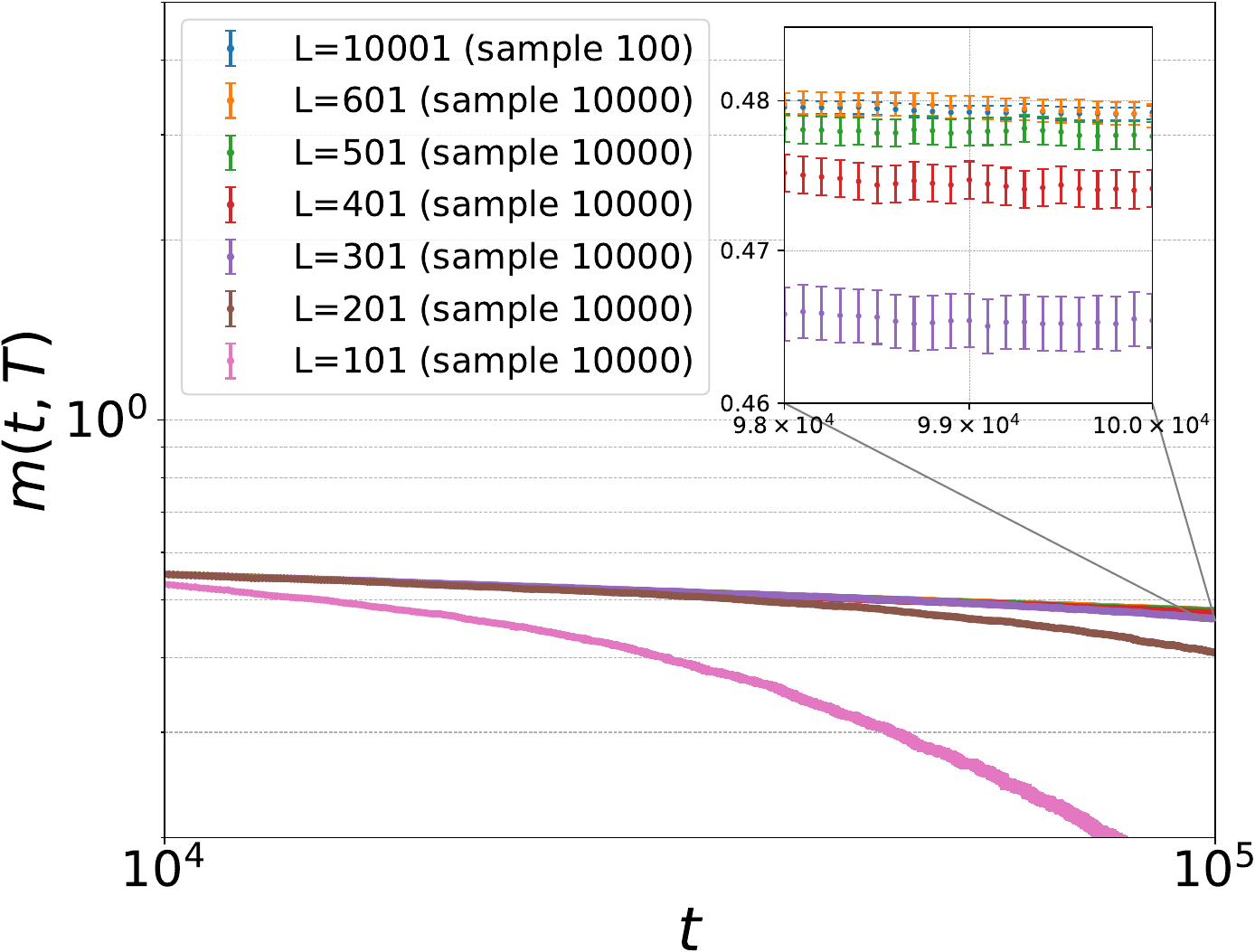}
    \end{overpic}
    
    \caption{Size dependence for the 2D 3-state Potts model. It shows that the dependence becomes negligible for $L \geq 601$. The data are plotted every 100 MCS. The legends indicate the number of samples used in each simulation. 
    }
    \label{fig:2d 3state potts size dependence}
\end{figure}

For a reliable dynamical scaling analysis, finite-size effects must be sufficiently small. The size dependence is examined in Fig.~\ref{fig:2d 3state potts size dependence}. From this result, we confirm that finite-size effects can be safely neglected up to $10^5$ MCS for $L \geq 601$. Based on this observation, we adopt $L=10001$ for the production runs and perform simulations up to $10^5$ MCS with 100 samples to construct the dataset.

From the resulting dataset, we exclude the initial relaxation regime that does not follow the dynamical scaling behavior, namely the data for $t<1000$ MCS, and use the remaining data for the final analysis. The total number of data points employed in this analysis is 2,475,025. Using this dataset, we perform the dynamical scaling analysis for the 2D 3-state Potts model. Following the results obtained for the 2D Ising model in Sec.~\ref{result:2D ising model}, the batch size is set to 256. The number of epochs is fixed to 10999, and convergence is confirmed from the behavior of the loss function and the physical parameters.

\begin{figure}[htbp]
    \centering
    \vspace{1.0em}
    \begin{overpic}[width=1.0\linewidth]{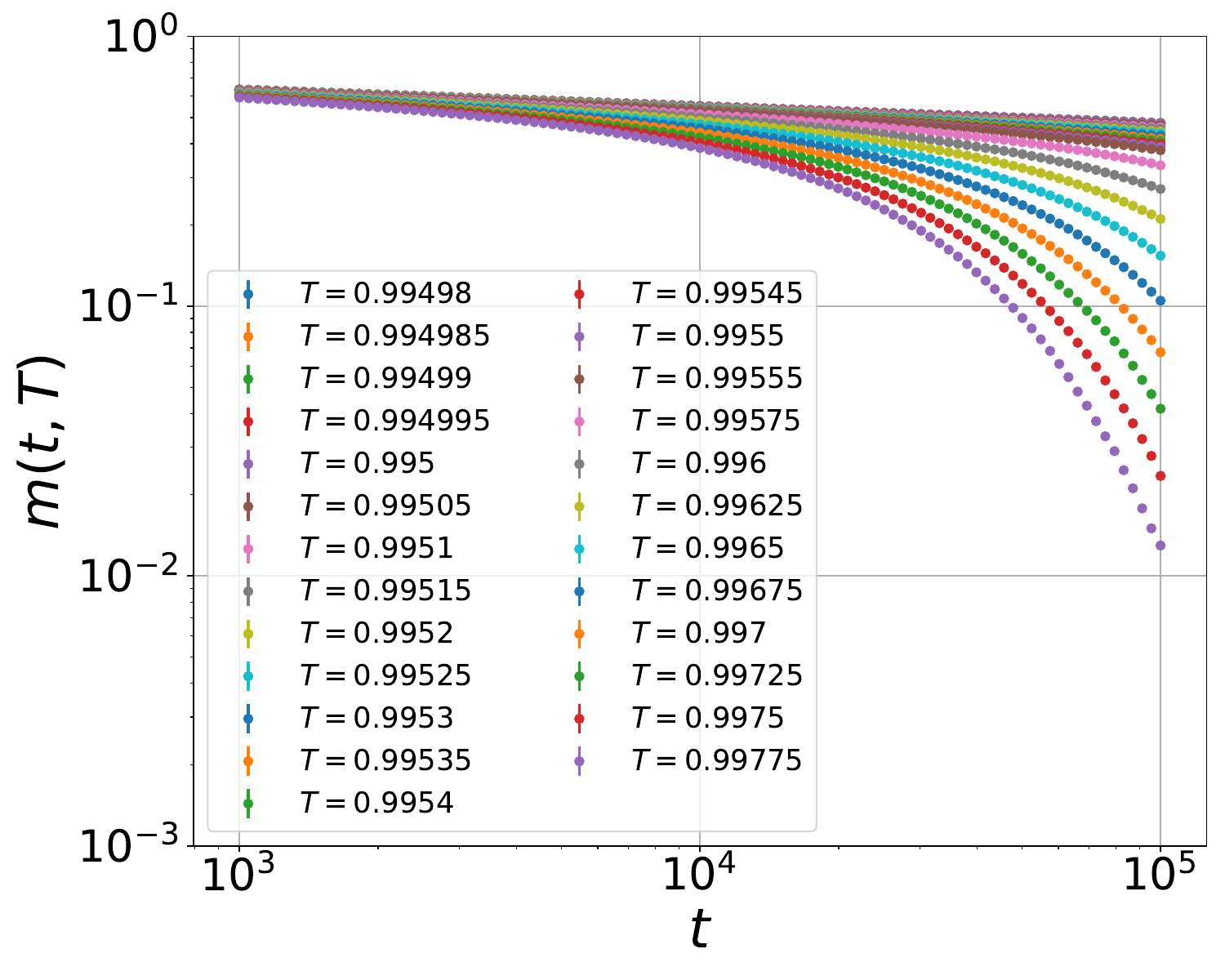}
        \put(90,80){\large (a)} 
    \end{overpic}
    \vspace{0.5em}
    \begin{overpic}[width=1.0\linewidth]{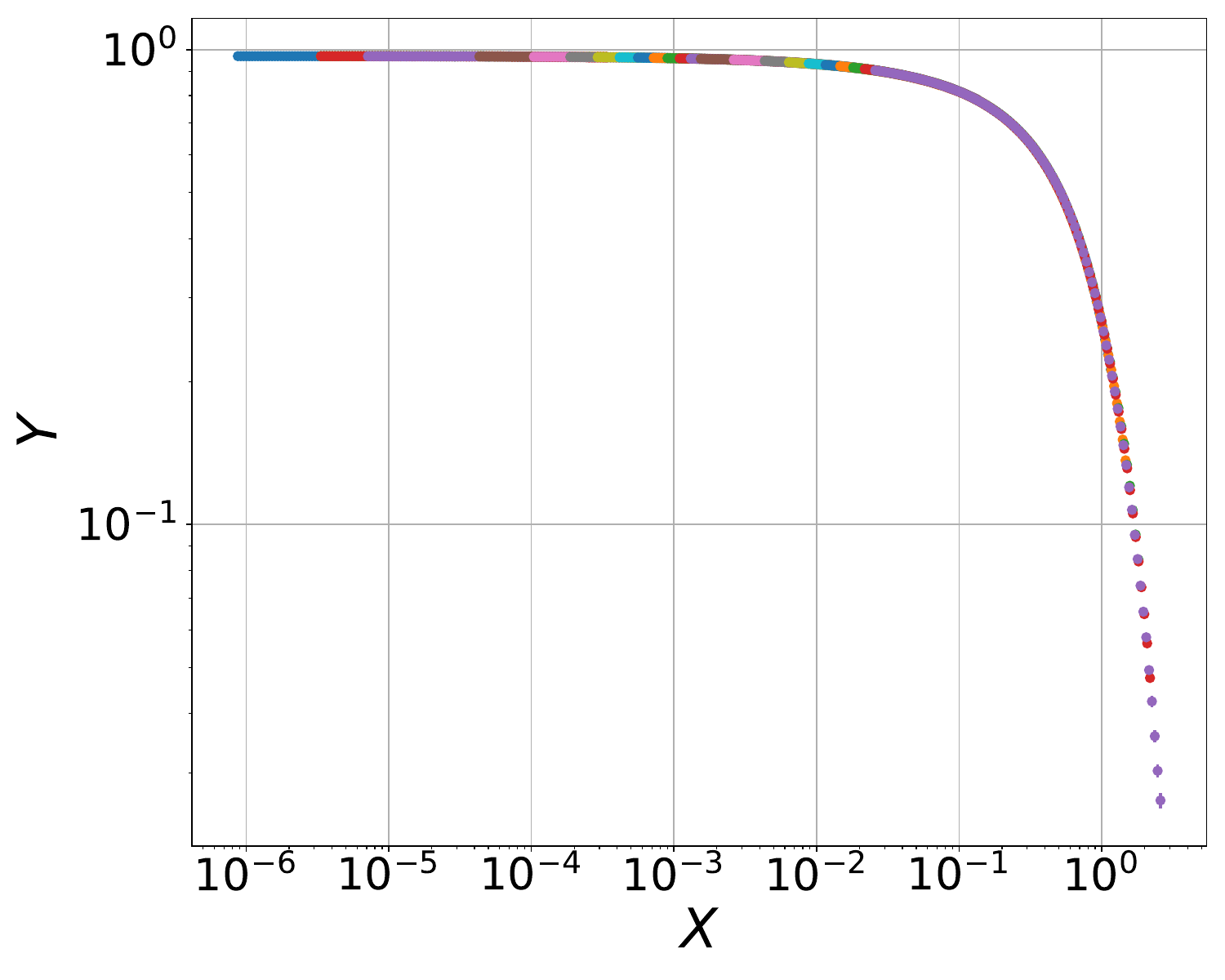}
        \put(90,80){\large (b)} 
    \end{overpic}
    \caption{Relaxation data and the corresponding dynamical scaling plot for the 2D 3-state Potts model analyzed in this study. In both panels, 101 data points were sampled for each temperature to visualize the data. (a) presents the relaxation behavior after eliminating the early-time regime, and these processed data were used for parameter estimation with the neural network. (b) displays the resulting scaling collapse. The obtained parameters are $T_{\mathrm{c}} = 0.994971(1)$, $\lambda = 0.0612(1)$, and $b = 1.795(1)$.}
    \label{fig:2d 3state potts scaling plot}
\end{figure}

The estimated parameters obtained under these conditions, together with those inferred by GPR, are summarized in Table~\ref{2d 3state potts result table}. The corresponding scaling plot based on the present method is shown in Fig.~\ref{fig:2d 3state potts scaling plot}. As in Sec.~\ref{result:2D ising model}, the error bars are evaluated by the bootstrap method with 128 resamplings. For the GPR analysis, 80 data points per temperature are extracted at equal intervals on a logarithmic time scale.

The obtained results eliminate the deviation observed in the GPR-based analysis by enabling the use of the full dataset, and are consistent with the exact critical temperature of the 2D 3-state Potts model within the error bars. 
Combined with the results for the 2D Ising model presented in Sec.~\ref{result:2D ising model}, these findings demonstrate that the improved computational efficiency of the present method leads to more reliable parameter estimation.

\begin{table}
\centering
\caption{Comparison of the estimated scaling parameters for the 2D 3-state Potts model.}
\label{2d 3state potts result table}
\setlength{\tabcolsep}{4pt} 
\begin{tabular}{clll}\hline\hline
Reference &  $T_{\mathrm{c}}$ & $\lambda$ & $b$ \\ \hline\hline
Exact Value & 0.9949728$\ldots$ & & \\ \hline
Neural Network & 0.994971(1) & 0.0612(1)  & 1.795(1) \\
GPR &  0.9949567(4) & 0.06060(2) & 1.8031(7) \\
\hline\hline
\end{tabular}
\end{table}

\section{Summary and Discussion}
\label{sec:Summary}
In this study, we propose a dynamical scaling analysis based on deep learning and demonstrate its validity by applying it to the 2D Ising model and the 2D 3-state Potts model, both of which exhibit a second-order phase transition with an exact critical temperature. 
Because the neural-network-based approach significantly reduces the computational cost compared with GPR, it enables the use of the full dataset in the scaling analysis. 
As a result, the estimated critical temperatures agree with the exact values and show improved accuracy compared with those obtained by the GPR-based analysis.
These findings establish the reliability of the proposed method as a general framework for dynamical scaling analysis.

The advantage of dynamical scaling analysis lies not only in its computational efficiency but also in its broad applicability. 
It can be applied to a variety of systems, including frustrated systems, systems exhibiting KT transitions, disordered systems, and percolation models~\cite{Hagiwara2022}. Extending the present approach to such systems remains a subject for future investigations.

Furthermore, the dynamical scaling method enables the estimation of physical quantities such as the correlation length in nonequilibrium processes\cite{Nakamura2016,Nakamura2019}. The application of the present framework to these types of problems also constitutes an important direction for future research.
In addition, the dynamical scaling method also allows for the estimation of critical exponents, and it has been shown to be particularly effective in the analysis of spin-glass systems~\cite{Yamamoto2004,Nakamura2006,Nakamura2016,Nakamura2019,Terasawa2023}.
However, since this method does not extrapolate the scaling regime as $t\rightarrow\infty$, the estimated values may be affected by finite-time corrections.
Recently, a systematic extrapolation method has been proposed to solve this issue\cite{Osada2024}.
Combining this technique with the dynamical scaling method may provide a way to overcome the limitation, which constitutes another important direction for future research.

The development of deep learning techniques aimed at achieving better performance has been particularly active in recent years. For instance, many studies have proposed optimizers that are expected to converge faster than Adam, which was employed in the present study~\cite{defazio2024roadscheduled,adamW}. Incorporating such methods may lead to further improvements in the proposed approach.

\section*{Acknowledgments}
 This work was supported by JST SPRING, Grant Number JPMJSP2131. The authors are also grateful to the Supercomputer Center at the Institute for Solid State Physics, University of Tokyo, for the use of their facilities.

\end{document}